\documentclass{elsart5}

\usepackage{graphicx}

\usepackage{amsmath}

\newcommand{\cgo}{CuGeO$_3$}
\newcommand{\sds}[2]{\vec{S}_{#1}\!\cdot\!\vec{S}_{#2}}

\begin{document}

\begin{frontmatter}

\title{The spin-Peierls chain revisited}

\author[aff1]{Georg Hager\corauthref{cor1}}
\ead{georg.hager@rrze.uni-erlangen.de}
\corauth[cor1]{}
\author[aff2]{Alexander Wei\ss{}e}
\author[aff1]{Gerhard Wellein}
\author[aff3]{Eric Jeckelmann}
\author[aff2]{Holger Fehske}
\address[aff1]{Regionales Rechenzentrum Erlangen, HPC Services, Martensstra{\ss}e 1, 91058 Erlangen, Germany}
\address[aff2]{Institut f{\"u}r Physik, Ernst-Moritz-Arndt-Universit{\"a}t Greifswald, Domstra{\ss}e 10a, 17487 Greifswald, Germany}
\address[aff3]{Institut f{\"u}r Theoretische Physik, Universit{\"a}t Hannover, Appelstra{\ss}e 2, 30167 Hannover, Germany}
\received{24 May 2006}
\revised{25 May 2006}
\accepted{26 May 2006}


\begin{abstract}
  We extend previous analytical studies of the ground-state phase
  diagram of a one-dimensional Heisenberg spin chain coupled to
  optical phonons, which for increasing spin-lattice coupling
  undergoes a quantum phase transition from a gap-less to a gaped
  phase with finite lattice dimerisation. We check the analytical
  results against established four-block and new two-block density
  matrix renormalisation group (DMRG) calculations.
  Different finite-size scaling behaviour of the spin excitation gaps
  is found in the adiabatic and anti-adiabatic regimes. 
\end{abstract}


\end{frontmatter}


Quantum phase transitions in low-dimensional correlated systems have
attracted considerable attention over the last decades. An example is
the Peierls instability in quasi-one-dimensional spin systems, i.e.,
the instability of a uniform spin chain towards dimerisation induced
by the interaction with lattice degrees of freedom. Starting in the
seventies with organic compounds~\cite{Mil83}, the interest in the
Peierls instability was renewed with the discovery of a spin-Peierls
(SP) transition in the inorganic compound \cgo{} in 1993 by
Hase~et~al.~\cite{HTU93}. The most significant feature distinguishing
\cgo{} from other SP-compounds is the high frequency $\omega$ of the
involved optical phonons, which is comparable to the magnetic exchange
interaction~$J$.

As an archetypal model for this type of SP system we consider the
antiferromagnetic Heisenberg chain coupled to harmonic Einstein
oscillators,
\begin{multline}\label{model}
  H = J\sum_i \sds{i}{i+1} + \omega\sum_i b_i^\dagger b_i \\
  + 
\bar{g} \sum_i (b_i^\dagger + b_i)(\sds{i}{i+1} - \sds{i}{i-1})\,,
\end{multline}
where $\vec{S}_{i}$ denote spin-$\frac{1}{2}$ operators at lattice site
$i$, and $b_i^{\dagger}$ and $b_i$ are phonon creation and
annihilation operators, respectively. This model has been studied with
a number of analytical and numerical methods, including second-order
flow equations~\cite{Uh98}, the combination of Schrieffer-Wolff
transformations~\cite{SW66} with a variational ansatz~\cite{WWF99}, as
well as exact diagonalisation~\cite{WWF99} and a four-block variant of
the density matrix renormalisation group (DRMG)~\cite{BMH99}. 
All these studies agree on the main finding that for finite phonon
frequency the system undergoes a transition from a phase with gapless
spin excitations to a dimerised phase with massive spin excitations
only for some finite value of the rescaled spin-phonon coupling 
$g=\bar{g}/\omega$. 

In the anti-adiabatic
limit $\omega\to\infty$ this critical coupling is approaching zero,
whereas in the adiabatic limit $\omega\to 0$ flow equations and DMRG
point towards a finite limiting value of $g_c$. This is equivalent to the
bare critical coupling $\bar{g_c}$ approaching zero linearly with $\omega$.
To second order in $g$, the Schrieffer-Wolff approach is
equivalent to the flow-equation result, but when fourth order terms
are included the critical coupling diverges for decreasing $\omega$.
This finding, of course, cast doubts on the method and on the
decoupling of spins and phonons by unitary transformations in general.

Using improved spin algebra codes in this article we therefore
reconsider the Schrieffer-Wolff approach for the model of Eq.~\eqref{model}.
In more detail, we try to decouple spin and phonon degrees of freedoms
by applying a unitary transformation $\tilde H = \exp(S) H \exp(-S)$
that removes interaction terms linear in~$g$,
\begin{equation}
  S = g \sum_i (b_i^\dagger - b_i)(\sds{i}{i+1} - \sds{i}{i-1})\,.
\end{equation}
Averaging the resulting Hamiltonian over the phonon vacuum we obtain
an effective spin model with long-ranged Heisenberg interactions. Such
frustrated spin chains are known to be susceptible to dimerisation,
and we can use the ratio of the next-nearest-neighbour to the
nearest-neighbour exchange, $\alpha_{\text{eff}}$, as an indicator of
the phase transition, which for the frustrated spin-$\frac{1}{2}$ 
chain
\begin{equation}\label{frust}
  H = \sum_i J (\sds{i}{i+1} + \alpha\sds{i}{i+2})
\end{equation}
occurs at $\alpha_c = 0.241167$ \cite{ON92*CCE95*Eg96}. The
transformed Hamiltonian $\tilde H$ can be expanded in a series of
iterated commutators,
  $\tilde H = \sum_k [S,H]_k / k!,$ 
where $[S,H]_{k+1} = [S,[S,H]_k]$ and $[S,H]_0 = H$. For orders $k>2$
the evaluation of these commutators rapidly becomes complicated, and
is feasible only with efficient computer algebra tools. We were now
able to push the limit of the expansion from order $4$ to order $8$.
Neglecting terms with more than two interacting spins, the resulting
effective spin Hamiltonian reads
\begin{equation}
  H_{\text{eff}} = J_0 N + \sum_i\sum_{n=1}^{5} J_n \sds{i}{i+n} 
\end{equation}
with $J_n = J\sum_{j=0}^{4} c_{j,n}\ g^{2j}$ and the coefficients
$c_{j,n}$ collected in Table~\ref{t.1}.

\begin{table}
  \caption{Expansion of the effective exchange interactions $J_n/J$ up to order $8$ in the electron phonon coupling $g$.}
  \label{t.1}
  \begin{center}
    \setlength{\tabcolsep}{1.5mm}
    \begin{tabular}{l|ccccc}
      & $ g^0 $ & $ g^2 $ & $ g^4 $ & $ g^6 $ & $ g^8$ \\
      \hline
      $J_0 $ &  
      & $ -\frac{3 \omega}{8} $ & $ \frac{9 \omega}{64} $ & $ -\frac{7 \omega}{64} $ & $ \frac{5 \omega}{64} $ \\
      $J_1 $ & $ 1 $ & $ \omega-\frac{3}{2} $ & $ -\frac{9 \omega}{8}+\frac{59}{24} $ & $ \frac{205 \omega}{144}-\frac{1123}{360} $ & $ -\frac{545 \omega}{384}+\frac{118791}{35840} $ \\
      $J_2 $ &   & $ \frac{\omega}{2}+\frac{3}{2} $ & $ \frac{\omega}{16}-\frac{25}{8} $ & $ -\frac{55 \omega}{144} +\frac{145}{32}$ & $ \frac{313 \omega}{512}-\frac{70573}{13440} $ \\
      $J_3 $ &   &   & $ \frac{5 \omega}{8}+\frac{2}{3} $ & $ -\frac{83 \omega}{96}-\frac{3023}{1920} $ & $ \frac{1255 \omega}{1536}+\frac{500191}{215040} $ \\
      $J_4 $ &   &   &   & $ \frac{31 \omega}{128}+\frac{937}{5760} $ & $ -\frac{2551 \omega}{5760}-\frac{44533}{107520} $ \\
      $J_5 $ &   &   &   &   & $ \frac{131 \omega}{2560}+\frac{5297}{215040}$
    \end{tabular}
  \end{center}
\end{table}

\begin{figure}
  \includegraphics[width=\linewidth,clip]{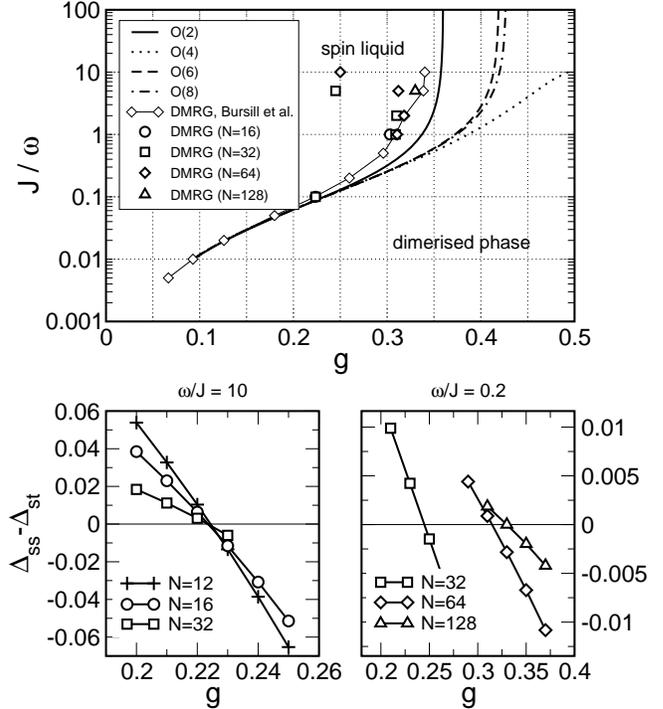}
  \caption{Upper panel: Ground-state phase diagram of the SP
    model, Eq.~\eqref{model}. Lower panel: System size dependence of
    the level crossing at $g_c$ in the adiabatic and anti-adiabatic
    limit.}\label{f.pd}
\end{figure}

In Figure~\ref{f.pd} we show the phase diagram obtained from the
condition $\alpha_{\text{eff}}=J_2/J_1=a_c$. The resulting critical
coupling $g_c$ oscillates with increasing order in $g$.  
Here the second and fourth correspond to the known results of Refs.~\cite{Uh98}
and~\cite{WWF99}, respectively, but convergence is achieved only when
going well beyond that.

It turns out, however, that the lowest order result shows the best
agreement with the four-block DMRG data of Bursill
et~al.~\cite{BMH99}. On the one hand, this may not seem surprising,
since the concept of integrating out phonon degrees of freedom usually
is appropriate only in the anti-adiabatic limit. On the other hand,
the smallest energy scale in the problem always corresponds to the
spin degrees of freedom, and therefore an effective spin model should
be able to describe the phase transition.

Numerically, the value of $g_c$ is obtained from the level crossing of
the lowest singlet and triplet excitations of the full spin-phonon
model, Eq.~\eqref{model}, i.e., from the same criterion applied to get
$\alpha_c$ of the frustrated spin chain, Eq.~\eqref{frust}. In the
adiabatic limit this procedure is rather delicate, since the
appropriate (finite-size) gaps must be smaller than
$\omega$, requiring large system sizes. In
addition, $g_c$ has a noticeable system size dependence, which is
negligible in the anti-adiabatic case (see lower panels of
Figure~\ref{f.pd}). To cross-check this numerical data we performed
new, large-scale parallel DMRG calculations~\cite{HJFW04} using the standard
lattice growth method of two sites per iteration. 
For comparable lattice sizes we find very good agreement.

How the unitary transformation scheme can be modified to yield 
improved effective Hamiltonians for the low energy physics 
of spin-Peierls systems will be the subject
of future studies.

We thank HLRN and NERSC for providing computational resources.



\begin{thebibliography}{10}
\expandafter\ifx\csname url\endcsname\relax
  \def\url#1{\texttt{#1}}\fi
\expandafter\ifx\csname urlprefix\endcsname\relax\def\urlprefix{URL }\fi

\bibitem{Mil83}
J.~S. Miller (Ed.), Extended Linear Chain Compounds, Vol.~3, Plenum, New York,
  1983.

\bibitem{HTU93}
M.~Hase, I.~Terasaki, K.~Uchinokura, Phys. Rev. Lett. 70 (1993) 3651.

\bibitem{Uh98}
G.~S. Uhrig, Phys. Rev. B 57 (1998) R14004.

\bibitem{SW66}
J.~R. Schrieffer, P.~A. Wolff, Phys. Rev. 149 (1966) 491.

\bibitem{WWF99}
A.~Wei{\ss}e, G.~Wellein, H.~Fehske, Phys. Rev. B 60 (1999) 6566.

\bibitem{BMH99}
R.~J. Bursill, R.~H. McKenzie, C.~J. Hamer, Phys. Rev. Lett. 83 (1999) 408.

\bibitem{ON92*CCE95*Eg96}
K.~Okamoto, K.~Nomura, Phys. Lett. A 169 (1992) 433;
G.~Castilla, S.~Chakravarty, V.~J. Emery, Phys. Rev. Lett. 75 (1995) 1823;
S.~Eggert, Phys. Rev. B 54 (1996) R9612.

\bibitem{HJFW04}
G.~Hager, E.~Jeckelmann, H.~Fehske, G.~Wellein, J. Comput. Phys. 194
  (2004) 795.

\end{thebibliography}

\end{document}